\documentclass[aps,prl,a4paper,reprint,nofootinbib,superscriptaddress,floatfix,preprintnumbers]{revtex4-1}

\usepackage{hyperref}
\usepackage{amsmath}
\usepackage{amsfonts}
\usepackage{amssymb}
\usepackage{color}
\usepackage[utf8]{inputenc}
\usepackage{graphicx}
\usepackage{microtype}
\usepackage{siunitx}
\usepackage{soul}

\newcommand{\header}[1]{\textbf{#1:}}
\newcommand{\Fermi}{\textit{Fermi}}

\begin{document}

\begin{flushleft}
LAPTH-050/18
\end{flushleft}

\title{Probing the \Fermi-LAT GeV excess with gravitational waves}
\date{\today}

\author{Francesca Calore}
\affiliation{Univ.~Grenoble Alpes, USMB, CNRS, LAPTh, F-74000 Annecy, France}
\author{Tania Regimbau}
\affiliation{LAPP, Univ.~Grenoble Alpes, USMB, CNRS/IN2P3, F-74000 Annecy, France}
\author{Pasquale Dario Serpico}
\affiliation{Univ.~Grenoble Alpes, USMB, CNRS, LAPTh, F-74000 Annecy, France}
\begin{abstract}
If the gamma-ray excess towards the inner Galaxy (GCE) detected in \Fermi-LAT data is due to millisecond pulsars (MSPs),
one expects an associated gravitational wave (GW) signal, whose intensity exceeds the disk MSP population emission by an order of magnitude. We compute the expected GW counterpart of the bulge MSP population based on fits of the GCE, and estimate the sensitivity reach of current and future terrestrial GW detectors. The bounds on the average population ellipticity $\varepsilon$ are competitive with the existing ones derived by LIGO/Virgo towards {\it known} MSPs. With a 10-year data taking in current LIGO/Virgo configuration, one would detect a signal at the level $\varepsilon\simeq 10^{-7}$, while $\varepsilon\simeq 10^{-8}$ would be attainable with a similar data taking period with a third generation GW detector. This sensitivity should be sufficient for crucial diagnostics on the GCE interpretation in terms of MSPs.  
\end{abstract}

\maketitle
\header{Introduction}
The detection of the binary neutron star merger GW170817 has ushered us in the era of multi-messenger astronomy with the use of gravitational waves (GWs)~\cite{TheLIGOScientific:2017qsa,GBM:2017lvd}. In the near future, the exploration and clarification of astrophysical events and processes via the combined use of different messengers is expected to become a routine tool for understanding  ``celestial labs''. 

Here, we focus on the perspectives of such an approach for elucidating the nature of the so-called Galactic Center Excess (GCE), inferred at GeV energies from several, independent analyses of \Fermi-LAT gamma-ray data collected from the inner part of our Galaxy (see e.g.~\cite{Gordon:2013vta,Calore:2014xka,TheFermi-LAT:2017vmf}). 
 
This intriguing signal roughly shares many of the properties (spectrum, intensity and morphology) expected for annihilation of weakly-interacting  dark matter particles, and has thus raised the interest of the astroparticle physics community in the last decade, see e.g.~\cite{Daylan:2014rsa,Calore:2014nla}. However, a number of arguments (clustering of the photon counts~\cite{Bartels:2015aea,Lee:2015fea}, tracing of the stellar mass in the bulge~\cite{Macias:2016nev,Bartels:2017vsx}, possible correlation with 511 keV emission~\cite{Bartels:2018eyb}) has been collected, suggesting rather an astrophysical origin of (most of) the excess emission. 
According to different analyses, the cumulative emission of point sources too faint to be detected individually by the LAT represents the dominant contribution to the GCE.   
If added to the similarity of the GCE spectrum to the one of gamma-ray millisecond pulsars (MSP)~\cite{Abazajian:2010zy}, a still unknown, sub-threshold population of MSPs in the Galactic bulge emerges as the leading culprit for the CGE. By itself, the observed GCE properties raise some interesting questions on the underlying progenitors of this new MSP population (see for instance the discussion in~\cite{Ploeg:2017vai,Eckner:2017oul}). However, alternative astrophysical scenarios (see e.g.~\cite{Petrovic:2014uda, Cholis:2015dea, Gaggero:2015nsa}) are still viable, in particular if contributing only partially to the GCE. The conjectured origin in an MSP population thus awaits confirmation from a source identification at some other wavelength, notably radio~\cite{Calore:2015bsx} or X-rays. 

Since MSPs are primary targets for terrestrial GW detectors, it is natural to wonder if the GW window may offer some complementary insights on the GCE nature. Hence our question: 
\textit{Under the hypothesis that the GCE is due to MSPs, what is the expected signal in GWs?}

In this Letter we explore what is the sensitivity of current and third generation GW detectors to a population of MSPs
in the Galactic bulge.
The spatial distribution and total number of MSPs are based on fits to the GCE properties, while the period distribution is taken from recent analyses of radio MSPs. We show that the prospects to probe this new population are very good and competitive with other, currently performed, GW searches.

\header{The expected GW signal from MSPs}
Neutron star high rotation velocities make any irregularity in their shape a quadrupolar source of GWs. This ``monochromatic'', continuous, emission is often considered the leading GW signal associated to pulsars (young and MSP alike)~\cite{1971ApJ...166..175I}, and the only one we consider in the following.
However, we note that the GW signal from spinning neutron stars could also be induced by other mechanisms~\cite{Andersson:1997xt, Owen:1998xg} which we neglect, but that would  further enhance the GW signal expected. In this respect, our estimate is conservative.

For a non-precessing triaxial body, rotating about its principal axis $z$, the mass quadrupole $Q$ is dominated by the $xx$ - $yy$ component of the moment of
inertia tensor, $I_{ij}$, and usually parameterized in terms of the ellipticity $\varepsilon$,
\begin{equation}
    \varepsilon\equiv \frac{I_{xx}-I_{yy}}{I_{zz}}\propto Q\,,
\end{equation}
where in practice $|\varepsilon|\ll 1$, and $I_{xx}\simeq I_{yy}\simeq I_{zz}=I$, for which a rather conservative value~\cite{Worley:2008cb} is $I = 1.1 \times 10^{38}$ kg m$^2$, as adopted in the following.
The GWs emitted by a star rotating with period $P$ and having ellipticity $\varepsilon$ have frequency $f=2/P$ and luminosity 
\begin{equation}
 L_{\rm GW}=\frac{2048\pi^6 G}{5 c^5}\frac{I^2  \varepsilon^2}{P^6}\,.
\end{equation}

Currently, the most sensitive search for GW emitted by {\it known} pulsars has been reported in~\cite{Abbott:2017ylp}, using O1 Advanced LIGO data. Single pulsar upper limits range from $\varepsilon<{\cal O}$(10$^{-1}$) to $\varepsilon<{\cal O}$(10$^{-8}$), with the bounds on most of the MSPs falling in the range between $10^{-7}$ and $10^{-6}$. While no detection has been reported, for relatively slow rotators, sometimes, these bounds surpass those inferred from spin-down arguments~\cite{Abbott:2017ylp}, although, for most MSPs, the typical limits from spin-down are actually about one order of magnitude stronger than current GW ones
~\cite{Abbott:2017ylp}. Interestingly enough, however, there are indications suggesting that MSPs have a minimum ellipticity $\varepsilon\gtrsim 10^{-9}$  (see~\cite{Woan:2018tey}), thus
implying that most MSPs spin down because of GW emission rather than magnetic braking.

Besides single source signals, the population of rotating neutron stars can contribute to the so-called stochastic GW background (SGWB)~\cite{Giazotto:1996ag, deFreitasPacheco:1996xt, Regimbau:2001kx}. The SGWB is believed to arise mostly from the superposition of GWs from a large number of unresolved sources.
The status of the art in current searches and sensitivities to the SGWB is reported in~\cite{TheLIGOScientific:2016dpb} and~\cite{TheLIGOScientific:2016xzw}, for an isotropic and 
directional background, respectively, both based on the LIGO run O1.
None of these searches has found evidence for a signal, and therefore they only set upper limits assuming a specific frequency spectrum.
As we shall argue, despite being ignored till now, we expect the MSP population in the Galactic bulge to be the {\it strongest} Galactic SGWB component in the LIGO/Virgo sensitivity band. Below, we estimate the current sensitivity to this signal and forecast the reach of future terrestrial detectors.

\header{GW from MSPs in the bulge}
Compared to other contributors to the SGWB, the one associated to the GCE is very anisotropic, peaking at the GC.
Our modelling of the MSP population in the Galactic bulge is based on gamma-ray results from analyses of the MSP source population~\cite{Bartels:2018xom, Ploeg:2017vai} and GCE~\cite{Calore:2014xka, Bartels:2017vsx}. The MSP spatial distribution matches the spatial properties of the GCE, namely it follows a power-law --with a slope of about 2.5-- with exponential cutoff at about 3 kpc~\cite{Calore:2014xka}.\footnote{Including the oblateness of the GCE profile, as found in~\cite{Bartels:2017vsx}, will have only a negligible impact on our estimate, given the poor angular resolution of GW detectors.}
The total number of sources is set by the GCE intensity and the MSP luminosity function. 
We adopt the most recent estimate of the GCE intensity (0.1 -- 100 GeV) from~\cite{Bartels:2017vsx}, $\mathcal{L}_{\rm GCE} = 2 \times 10^{37}$ erg/s. As for the MSP gamma-ray luminosity function, 
this has been derived recently from detected MSPs in the Galactic disk~\cite{Bartels:2018xom, Ploeg:2017vai}. Under the hypothesis that the GCE emission is all due to bulge MSPs and that the gamma-ray luminosity function of bulge MSPs is the same as the one of disk MSPs, we find that in total there are about $3.5 \times 10^4$ MSPs in the bulge, when adopting the best-fit parameterisation (i.e.~broken power-law) of the luminosity function from~\cite{Bartels:2018xom}.\footnote{For the best-fit parameters of a log-normal distribution and $L_{\rm min} > 10^{30} \rm erg/s$, we find instead $\log_{10}{N^{\rm b}_{\rm tot}} = 4.2$. We also mention that Ref.~\cite{Ploeg:2017vai} estimates $\log_{10}{N^{\rm b}_{\rm tot}}= 4.6$, above $L_{\rm min} > 10^{32} \rm erg/s$ for the GCE intensity from~\cite{Macias:2016nev}.} 
Typically, the gamma-ray luminosity function is especially uncertain at low luminosities and therefore the total number of sources can depend on the minimal luminosity assumed, $L_{\rm min}$. However, for the actual problem at hand the total number of sources varies very mildly with $L_{\rm min}$.
We note that an alternative way to estimate the number of MSPs in the bulge is to use instead the radio luminosity function of MSPs in globular clusters. Following Ref.~\cite{Calore:2015bsx}, the total numbers of MSPs expected in the bulge would be about 10$^4$, hence of the same order of magnitude.

Given the poor angular resolution of LIGO/Virgo, the details of the spatial distribution of the sources are irrelevant for the prediction of the GW signal, which thus results almost 
indistinguishable from a single source located at the GC with a characteristic frequency spectrum. In what follows, we assume a semi-aperture of the pointing cone of $\theta = 10^\circ$. A posteriori, this is a slightly conservative estimate of the angular resolution (see below). Moreover, we note that integrating over a larger $\theta$ would imply a moderately larger signal (see Appendix), thus making the estimates below conservative also in this respect. 

From the definition of the GW power spectral density $H(f)$ (see e.g.~\cite{Talukder:2014eba}\footnote{The additional factor of 2 that we include in Eq.~\ref{eq:Hf}, with respect to Ref.~\cite{Talukder:2014eba}, is required by the sum over the two GW polarisations.}), we write: 
\begin{equation}
    H_{\theta} (f) = \frac{32 \pi^4 G^2}{5 \, c^8}  \varepsilon^2  I^2  f^4   \mathcal{P}(f)   \int_{\rm l.o.s.} \frac{\mathcal{N}_{\theta}(s)}{s^2} ds \, . 
    \label{eq:Hf}
\end{equation}
$\mathcal{N}_{\theta}$ is the differential number of sources at a line of sight (l.o.s.) distance $s$ within a cone of semi-aperture $\theta$ around the GC. $\mathcal{P}(f)$ is the probability density function of MSP frequencies. 
As for $\varepsilon,\, I$, Eq.~\ref{eq:Hf} is valid under the assumption that they do not depend on frequency and position of the sources.
If they are not constant, the parameters $\varepsilon,\, I$ have to be intended as appropriate averages over the population in the sightlines and angular region probed. 
For illustrative purposes, we also define the number
\begin{equation}
N_{\theta}\equiv d_{\rm GC}^2  \int_{\rm l.o.s.} \frac{\mathcal{N}_{\theta}(s)}{s^2} ds\,,\label{Ntheta}
\end{equation}
which coincides with  the total number of sources in the limit in which they are all concentrated at the GC (point-like approximation), i.e. if:
\begin{equation}
\mathcal{N}_{\theta}(s)=N_\theta\delta(s-d_{\rm GC})\, .
    \label{eq:pointlike}
\end{equation}
Below, we set the GC distance to $d_{\rm GC}=8.5\,$kpc. Note that
 $H(f)$ is given per unit frequency, Hz$^{-1}$, and it is linearly proportional to the effective number of sources, $N_{\theta}$, and to the squared ellipticity, $ \varepsilon^2$.

\begin{figure}[ht!]
	\includegraphics[width=0.85\columnwidth]{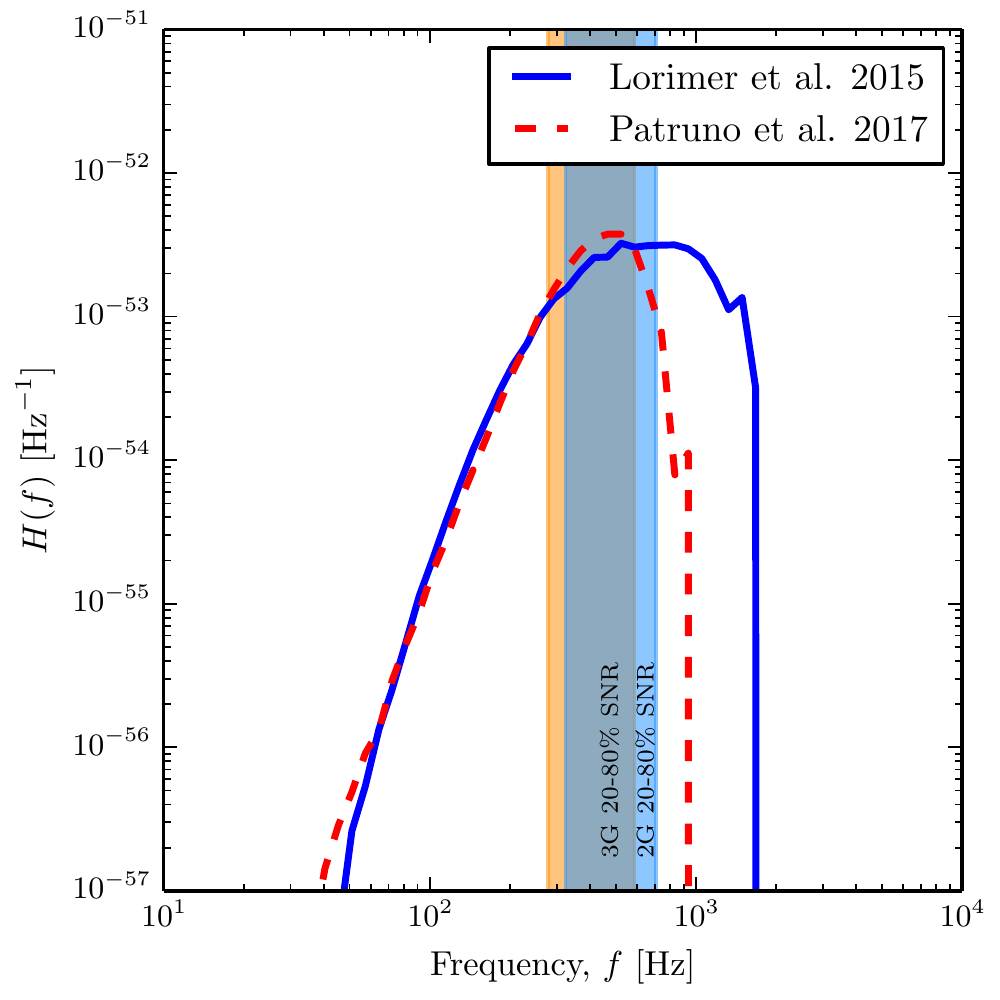}
    \caption{GW power spectral density $H(f)$ (see Eq.~\ref{eq:Hf}) for the MSP bulge population able to explain the GCE intensity, with $N_{\theta} = 10^{4}$ and assuming $\varepsilon = 10^{-7}$. Two period distributions are considered from~\cite{Lorimer:2015iga} (solid blue line) and from~\cite{Patruno:2017oum} (dashed red line). The rightmost and leftmost bands show the frequency ranges contributing to the central 60\% of the total signal-to-noise ratio ($SNR$) for the current (2G) and future (3G) generation of terrestrial GW detectors, respectively, for the log-normal period distribution~\cite{Lorimer:2015iga}.}
    \label{fig:Hf}
\end{figure}

As for the period distribution, which determines the functional dependence of $H(f)$ in Eq.~\ref{eq:Hf}, we adopt two possible parameterizations: The first is a log-normal distribution from Ref.~\cite{Lorimer:2015iga}, obtained by a fit to radio MSPs; the second is a Weibull distribution from a more recent statistical analysis of spin frequencies of an extended sample of radio MSPs~\cite{Patruno:2017oum}. The corresponding $H(f)$ are displayed in Fig.~\ref{fig:Hf}, where we assume $\varepsilon = 10^{-7}$, consistently with current upper limits for most of the pulsars analyzed in Ref.~\cite{Abbott:2017ylp}, and $N_\theta = 10^4$.

Till now, we have {\it assumed} that the bulge MSP component dominates with respect to the MSP disk population. However, we have checked that this is  indeed the case, when integrating the total number of MSPs (disk and bulge) along the sightline. To model disk MSPs we adopt the best-fit model of Ref.~\cite{Bartels:2018xom}, namely a Lorimer-disk spatial distribution, a broken power-law luminosity function, and the same period distribution as for bulge MSPs. The total number of disk MSPs is about $2.4\times 10^4$, 3500 of which contained in a cone of 10$^\circ$ semi-aperture towards the GC direction. In the direction of the GC,  disk MSPs amount to about 10\%  of bulge MSPs. (For a more detailed description with further quantitative results, we refer the interested reader to the Appendix.)
Although the  bulge signal dominance over the disk one has been illustrated for a specific model, we believe that it is valid in general, for any viable interpretation of the GCE in terms of unresolved MSPs in the bulge.  
 Ultimately, our conclusion relies on the empirical argument that---contrarily to the GCE---there is no evidence, yet, for a ``Galactic Disk Excess" in the gamma-ray band that could be attributed to MSPs, see in particular A.3.3 in~\cite{Bartels:2017vsx}.

\header{Sensitivity of LIGO/Virgo}
The methods used to search for the SGWB make minimal assumptions on the morphology of the signal, relying instead on excess coherence in the cross-correlated data streams from multiple detectors, as opposed to the independent and uncorrelated individual detector noise. To probe anisotropies in the SGWB, the standard method is the GW radiometer which is analogous to aperture synthesis used in radio astronomy~\cite{Ballmer:2005uw}. By applying appropriate time-varying delays between detectors it is possible to map the angular power distribution in a pixel or spherical harmonic basis. 

\begin{figure}[ht!]
	\includegraphics[width=0.85\columnwidth]{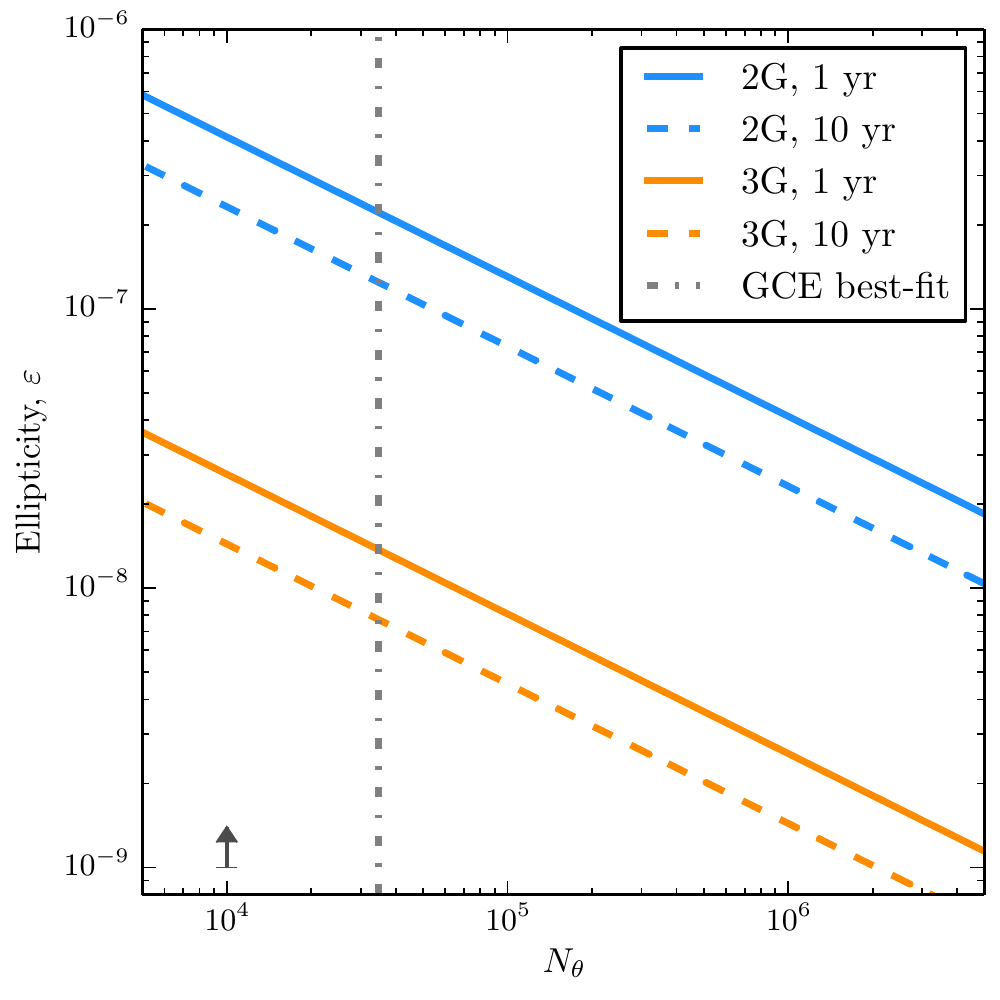}
    \caption{3$\sigma$ sensitivity of current (2G) and future (3G) generations of terrestrial GW detectors to a population of MSPs in the Galactic bulge.
     The vertical dot-dashed line is our total number of MSPs in the bulge required by fits to the GCE (based on~\cite{Bartels:2017vsx}), in the point-like approximation of Eq.~\ref{eq:pointlike}. The
     arrow indicates the lower limit on $\varepsilon$  according to~\cite{Woan:2018tey}.}
    \label{fig:Ntoteps}
\end{figure}

The detectability for a pair of detectors with given noise power spectral densities scales with $\varepsilon^2$ and depends on the observation time $T$ as $\sqrt{T}$ (see Eq.~16 of ~\cite{Ballmer:2005uw}).
Using the $H(f)$ calculated in the previous section and assuming the parameterization of~\cite{Lorimer:2015iga}, we obtain
\begin{equation}
SNR\simeq 0.18\,\{46\}\,\frac{N_\theta}{10^4}\left(\frac{\epsilon}{10^{-7}}\right)^2\sqrt{\frac{T}{1\,{\rm yr}}}\,,\label{scaling}
\end{equation}
where the first number refers to a network composed of the two LIGO detectors at Hanford and Livingston and Virgo, all at their designed sensitivity (hereafter 2G detectors);  the number in curly brackets corresponds to third generation detectors (3G), such as Einstein telescope~\cite{Punturo:2010zz} or Cosmic Explorer~\cite{Evans:2016mbw}. In particular, in Eq.~\ref{scaling} we assume two Cosmic Explorer detectors at the actual LIGO sites and one Einstein Telescope at the actual Virgo site. In Fig.~\ref{fig:Ntoteps}, we show the $3 \sigma$ sensitivity (i.e. $SNR=3$) of current (2G, top blue curves) and future (3G, bottom orange curves) GW detectors vs. $N_\theta$, for 1 year (solid curves) or 10 years (dashed curves).  For the benchmark value of $N_\theta$, corresponding to the total number of MSP in the bulge in the approximation of Eq.~\ref{eq:pointlike} and illustrated with a vertical dashed line, we obtain that
a signal from bulge MSPs with  $\varepsilon \simeq 2.2 (1.2)\times 10^{-7}$ is observable at $3 \sigma$ after one year (ten years) at 2G, while at 3G in the same time one attains $\varepsilon \simeq 1.4 (0.8) \times 10^{-8}$, respectively.  We stress that the search discussed here is directly sensitive to a population average of $\varepsilon^2$, contrarily to single source searches.

Concerning the sensitivity to MSP population models, the rightmost and leftmost vertical bands in Fig.~\ref{fig:Hf} show the frequency ranges $[f_{20}; f_{80}]$ contributing to the central 60\% (from 20\% to 80\%) of the $SNR$ for the 2G and 3G detectors, respectively, for the case~\cite{Lorimer:2015iga}. Namely, all $f<f_{20}$ contribute to 20\% of the $SNR$, and all $f>f_{80}$ to another 20\%. If we were to use instead the parameterization of~\cite{Patruno:2017oum}, the $SNR$ would only drop by 10\% (5\%) for 2G (3G) detectors, with a narrowing of the $[f_{20}; f_{80}]$ band by  $\sim$60\%, essentially due to a lower $f_{80}$ caused by the predicted drop of $H(f)$ at high frequency. 
Note that the most sensitive frequencies are around 400 Hz, hence we expect the diffraction-limited spot size on the sky to be $\simeq 7^\circ\lesssim 10^\circ$, see~\cite{TheLIGOScientific:2016xzw}, consistently with the angular scale considered here.

\header{Discussion and conclusions}
Motivated by the gamma-ray indications for the existence of a population of unresolved sources --most likely MSPs-- in the Galactic bulge~\cite{Bartels:2015aea, Lee:2015fea, Bartels:2017vsx, Macias:2016nev}, we have shown that the collective GW signal generated by $\mathcal{O}(10^4)$ MSPs at the GC would indeed constitute the dominant ``Galactic diffuse'' background due to MSPs, as opposed to the disk population conventionally searched for.
{\it This newly discussed target would thus represent the leading Galactic SGWB component in the LIGO/Virgo sensitivity region,} and as such should be considered as a benchmark of primary interest in future studies. The possible discovery of this signal would also provide an independent evidence that the GCE arises from unresolved point sources, rather than from dark matter particle annihilation. 
 The sensitivity to such a signal in the current generation of detectors (2G), which we estimate at $\varepsilon\sim 2\times 10^{-7}$ in one year, is already comparable to the {\it targeted searches} towards known MSPs reported in O1~\cite{Abbott:2017ylp}. With detectors of the third generation, one may improve the sensitivity to $\varepsilon$ by more than one order of magnitude.
 If the ellipticities are  comparable to current bounds from spin-down, one may  expect a GW discovery in 3G detectors. A null detection in 3G would support a lower value for $\varepsilon$, maybe close to the lower limit of $\varepsilon \gtrsim 10^{-9}$ derived in~\cite{Woan:2018tey}.

Within the same time frame, targeted studies towards nearby MSPs (such as PSR J1643-1224 and PSR J0711-6830) should lead to a detection at least in 3G~\cite{Woan:2018tey}: Our study thus opens the perspective of an interesting interplay between the GW knowledge from specific nearby disk MSPs and studies of the putative bulge MSP population, {\it even for a null result for the latter}. If the first individual MSPs will be identified in GW in the current 2G run, hence around $\varepsilon\gtrsim 10^{-8}$, there are good perspectives that the 3G may detect the unresolved bulge contribution as well. On the other hand, if the latter is not revealed, one might have to revise the hypothesis that the bulge population has the same period distribution as the disk one\footnote{We stress, in fact, that the inference of an MSP origin for the GCE (as opposed to a new population of pulsars with its own characteristics, or an alternative class of astrophysical objects and processes) is only circumstantial. Lacking single source identifications and period measurements, we adopted period estimates taken from the disk MSP population to gauge the GW signal level associated to the GCE. }, or  the currently favored interpretation of the GCE in terms of MSPs altogether. 
If no GW detection of single nearby MSP will be announced in the 2G run~\cite{Woan:2018tey}, the average ellipticity will be likely lower than 10$^{-8}$ (see Fig.3 of~\cite{Abbott:2017ylp}) and therefore we expect that no detection of the bulge population will take place at 3G either. In that case, it is the eventual {\it detection} of a bulge signal which would profoundly shake the foundation of the current MSP interpretation of the GCE. While speculative, this possibility is not necessarily far-fetched: here, for instance, we have neglected any contribution from a {\it young} pulsar population in the bulge.

At the same time, MSP radio searches are ongoing and expected to dig into the bulge population in the next few years with the operation of the MeerKAT (and later SKA) telescope. Of course, a discovery of bulge pulsars at radio frequencies and the measure of their period and spatial distribution will motivate a revisitation of the current calculations and strengthen the rationale for a dedicated GW pulsar search towards the inner Galaxy.

\vskip 5mm

\header{Acknowledgments}  
We thank C. Weniger and G. Zaharijas for comments on the manuscript, and  T.~D.~P.~Edwards for fruitful discussions.  
FC and PDS acknowledge support from Agence Nationale de la Recherche under the contract ANR-15-IDEX-02, project ``Unveiling the Galactic centre mystery'',
GCEM (PI: F. Calore).
\bibliography{GCE_GW}

\newpage
\section{Appendix}
\begin{figure}[!ht]
	\includegraphics[width=0.85\columnwidth]{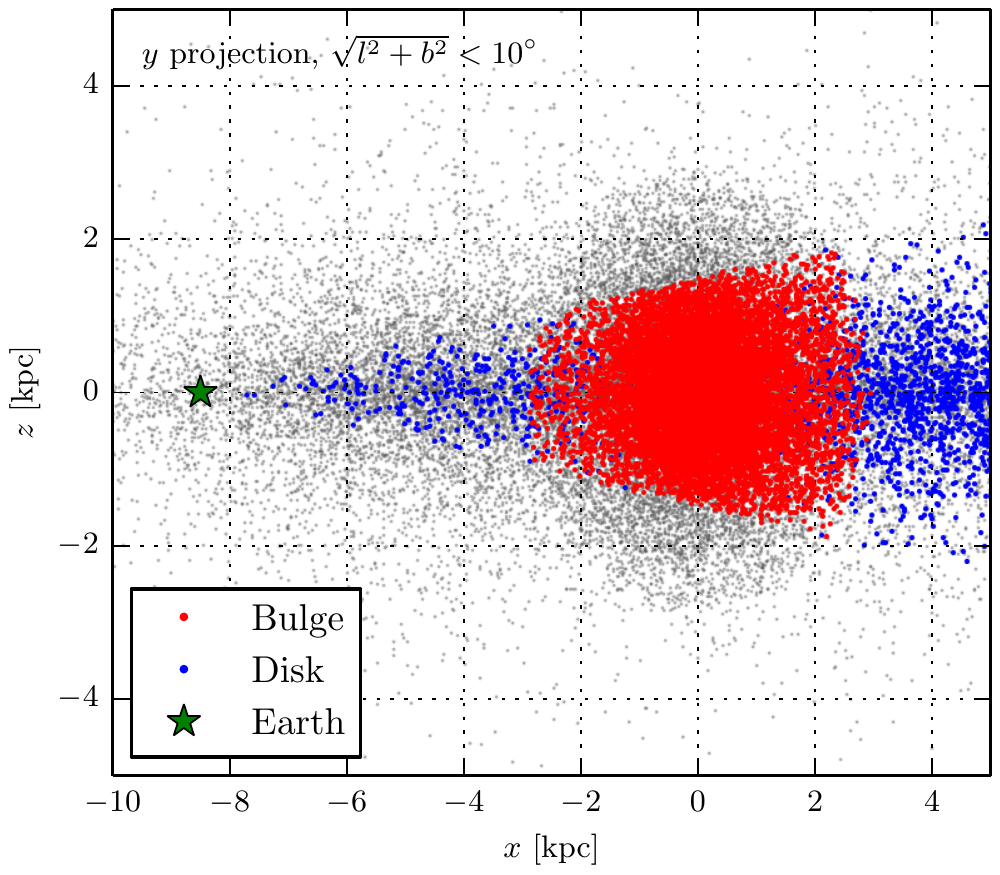}
    \caption{Vertical/lateral cross-section of the simulation of the Galaxy with red (blue) dots representing the bulge (disk) MSPs fulfilling the angular cut $\theta=\sqrt{l^2+b^2}<10^\circ$, while grey dots the MSPs outside such a geometrical cut  }
    \label{fig:MCsnapshot}
\end{figure}

\header{Contribution of disk MSPs}
To model the two MSP populations in the Galaxy, one in the bulge and the other in the disk, we assume that the MSP emission is universal and, therefore, that the two populations share the same luminosity function, period distribution, average ellipticity and moment of inertia, but have distinct spatial distributions. This means that $\mathcal{N}_{\theta}(s)$ in Eq.~\ref{eq:Hf} writes as $\mathcal{N}_{\theta}(s)=\mathcal{N}_{\theta}^b(s)+\mathcal{N}_{\theta}^d(s)$, and $N_\theta$ in Eq.~\ref{Ntheta} writes as
 $N_{\theta}=N_{\theta}^b+N_{\theta}^d$. These quantities are computed directly from a Monte Carlo simulation of the spatial distribution of MSP sources in the Galaxy, cut at the angle  $\theta$ with respect to the GC direction (as seen from Earth), and binned as a function of distance along the line of sight. In Fig.~\ref{fig:MCsnapshot} we show a cross-section of this simulation in the plane orthogonal to the Galactic disk and containing the Earth-GC line, with red (blue) dots representing the bulge (disk) MSPs fulfilling the geometrical constraint, while grey dots the MSPs outside the angular cut.

\begin{figure}[!ht]
	\includegraphics[width=0.85\columnwidth]{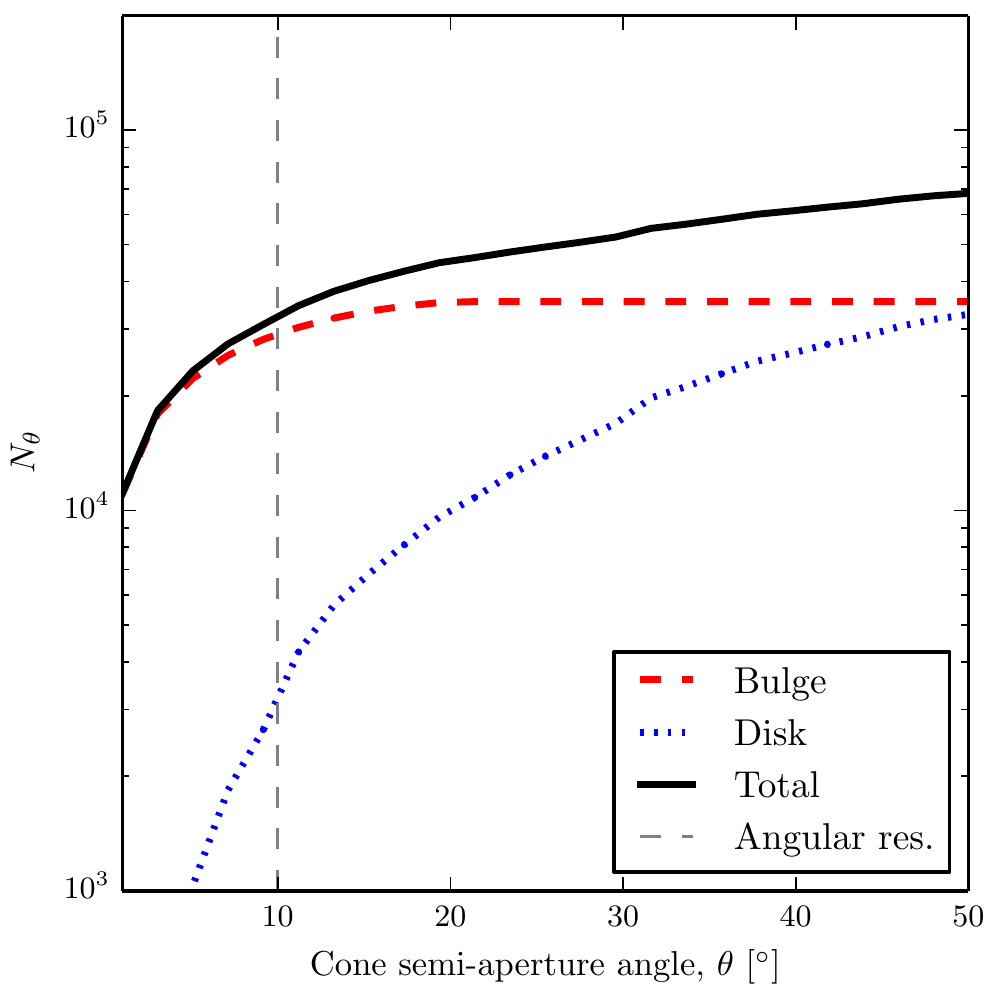}
    \caption{$N_{\theta}$ (Total), $N_{\theta}^b$ (Bulge), $N_{\theta}^d$ (Disk) as a function of the cone semi-aperture angle $\theta$ in black solid, red dashed, and blue dotted lines, respectively.  The dashed grey line indicates the benchmark value adopted for  the angular resolution of GW detectors considered, i.e. $\theta=10^\circ$.}
    \label{fig:bulgevsdisk}
\end{figure}

In Fig.~\ref{fig:bulgevsdisk}, we show the effective number, $N_{\theta}$, proportional to signal strength, as a function of $\theta$. The plot clearly illustrates that the total signal is dominated by $N_{\theta}^b$ (bulge MSPs) up to very large values of $\theta\simeq 50^\circ$. Also, note that $N_{\theta}^b$ almost saturates beyond $\theta\simeq 10^\circ$, so that taking this angle as benchmark, while conservative, is not expected to lead to a large underestimate of the signal.

\begin{figure}[ht]
	\includegraphics[width=0.85\columnwidth]{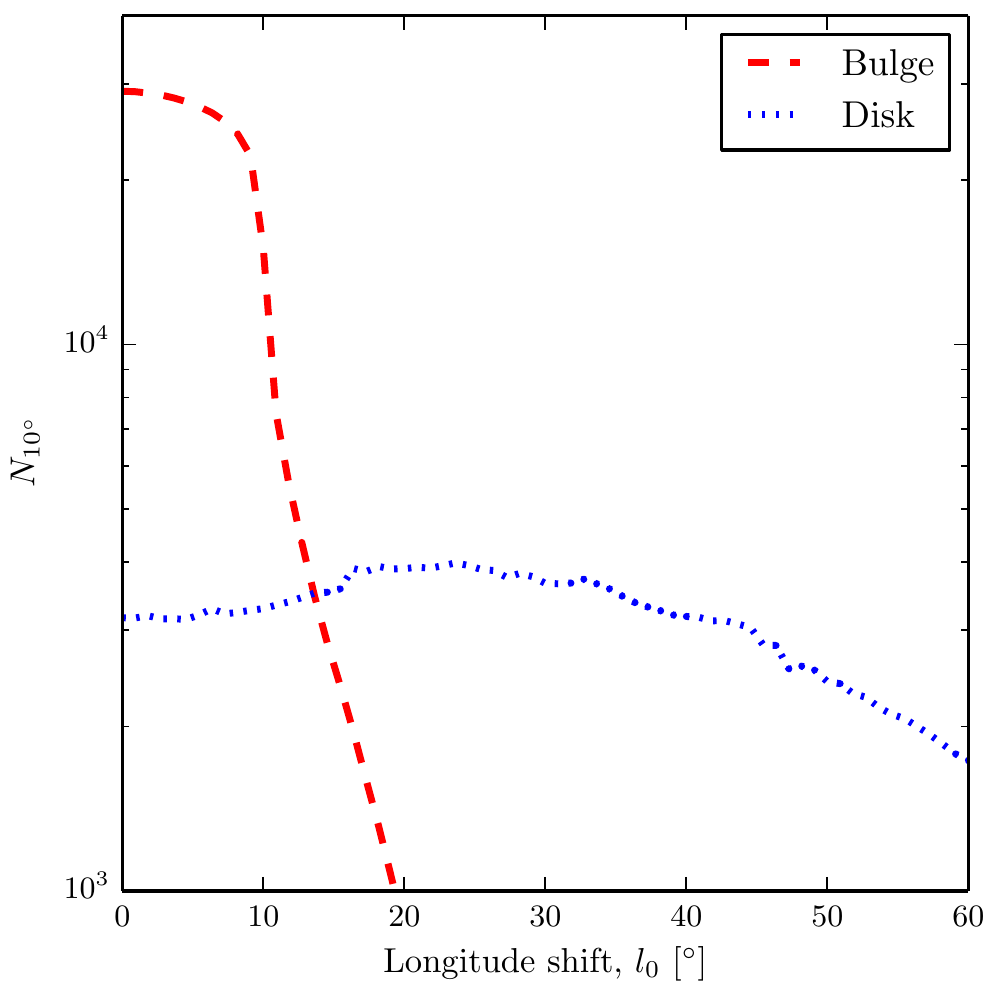}
    \caption{$N_{10^\circ}$'s  (i.e. within a cone of 10$^\circ$ semi-aperture) for the bulge (red dashed line) and disk (blue dotted line) populations as a function of the longitude shift $l_0$ of the center of the cone (all set at zero Galactic latitude). The behaviour is symmetric in $l$.}
    \label{fig:disk}
\end{figure}

We also check that the GW signal induced by the MSP disk population along the Galactic plane --i.e. centering the cone at different longitudes, $l_0$-- is always subdominant with respect to the signal from the GC direction, see Fig.~\ref{fig:disk}. This implies that the ``point-like source'' analysis applied here  is maximally sensitive to the bulge MSP population.

The MSP disk population thus represents a sub-dominant contribution to both the unresolved gamma-ray and GW signals. At the same time, the two populations represent ideal targets for rather complementary search techniques, with the disk one more appropriately probed via anisotropic, SGWB methods~\cite{ongoing}.

\end{document}